# Status report of the ANTARES project

Maurizio Spurio, *Department of Physics and INFN- 40100 Bologna (Italy) –email:* spurio@bo.infn.it,
for the ANTARES Collaboration

*Abstract*—The detection of very high energy neutrinos of galactic/extragalactic origin requires very large detectors and a large overburden as a shield against the background of cosmic ray muons. ANTARES is at present the largest (effective area ~0.05 km$^2$) experiment currently under construction in the northern hemisphere. It is being built and installed at a depth of 2500m in the Mediterranean sea, near the Southern French coast, by a large European collaboration. A three-dimensional array of photomultipliers are used to detect the Cherenkov light emitted by neutrino-induced muons. The array, when completed, will consists of 12 lines each covering a vertical length of about 480 m and equipped with 75 photomultipliers arranged in triplets. The readout electronics is connected to an on-shore laboratory through a 42 km long electro-optical cable. The final detector design has been completed. An instrumented line (called MILOM) has been installed in the spring of 2005; the first string (Line 1) is in acquisition starting from February 2006, and the second (Line 2) from September 2006. The physics motivations of the experiment, the details of the construction and installation, together with preliminary results obtained using the MILOM and Line 1 are presented.

## I. INTRODUCTION

A promising challenge for exploring the Universe is the detection of high-energy neutrinos. Fluxes of high energy neutrinos are expected from sources like supernova remnants [1], micro-quasars [2], active galactic nuclei [3] or gamma ray bursters [4]. Protons are accelerated in these sources up to ultra high energies. In the *astrophysical beam dump* model, hadrons are produced in the interaction of those protons with other protons (or nuclei) or low energy photons. Cascades induced by the produced hadrons lead to a flux of secondary mesons which could decay to very high energy neutrinos. Neutrinos escape from the source and travel large distance to the Earth. The weakly interacting nature of neutrinos, and the fact that they are not deflected by magnetic fields, make them unique probes of the regions of the Universe at distances larger than 50 Mpc. Another source of neutrinos (in the energy range $100 \leq E_\nu \leq 1000\, GeV$ ) could be the annihilation of the lightest super-symmetric particles, the neutralinos, trapped (by the combined effect of elastic scattering and gravitation) in celestial bodies like the Sun or the Galactic Centre. If the very low interaction cross section of neutrinos with matter is an advantage for exploring remote regions of the universe, it constitutes a drawback for their detection. Considering the expected fluxes, a huge apparatus is necessary to detect a few neutrino events per year. It is commonly assumed that the required detector will extend over 1 km$^3$. Moreover, the detector must be protected against the intense flux of cosmic rays. Natural Cherenkov radiators (water or ice) provide a large active volume at reasonable costs. The indirect detection of neutrinos through muons profits from the increase of the target region related to the muon range. The Cherenkov light emitted by charged particles is detected by an array of photomultiplier tubes (PMTs) which are housed in a pressure-resistant glass sphere. The muon direction and energy are measured using the arrival times and amplitudes of the PMTs.

Experiments (BAIKAL [5], AMANDA [6]) have been built over the last decades, and others (NESTOR [7], NEMO [8], ICECUBE [9]) are being designed or constructed for the detection of very high energy neutrinos of cosmic origin. No events have been detected so far above the unavoidable background of atmospheric neutrinos. The ANTARES experiment, that will here be discussed, is a medium-sized experiment (the effective area is $\cong 0.05$ km$^2$, of the same order of magnitude as the one of AMANDA) now being assembled in the Mediterranean, off the southern French coast.

## II. THE ANTARES DESIGN

The ANTARES (Astronomy with a Neutrino Telecope and Abyss environmental RESearch) project started in 1996 and involves physicists, astronomers, sea science experts and engineers from France, Germany, Italy, Russia, Spain and The Netherlands. The detector is being deployed at a depth of about 2500 m in the Mediterranean sea, 37 km off-shore of La Seyne-sur-Mer, near Toulon (France). ANTARES will look at the whole southern sky hemisphere, and a significant fraction of the northern one. The Galactic Centre will be visible for 67% of the day. It complementary the apparatus located at the South Pole, with a sky overlap of around 1.5π sr.

An intense R&D and site evaluation program has provided the relevant environmental parameters of the detector site [10,11,12]. Extensive surveys of the water optical properties have been carried out, along with a detailed analysis of the optical background due to the water $^{40}$K, bioluminescence, biofouling on optical modules and light transmission properties [13,14].

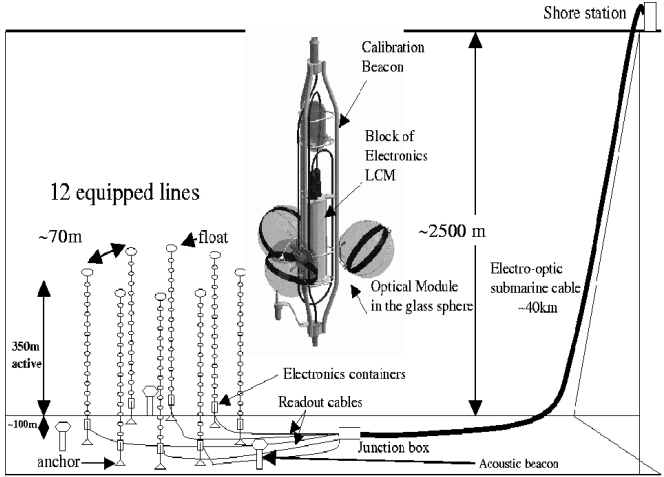

Fig. 1. Schematic view of the ANTARES detector.

The detector, illustrated in figure 1, consists of 12 lines made of mechanically resistant electro-optical cables. Each line will be anchored at the sea bed at distances of about 70 m one from each other, and tensioned by buoys at the top. Each line has a total length of about 450 m, and is equipped with 75 optical modules arranged in triplets (*storey*) and oriented $120^o$ in azimuth apart from each other, and $45^o$ below the equator. The vertical distance between adjacent storeys is 14.5 m. Each line is connected to a Junction Box (deployed in 2002), which transmits power and sends the data to the ground station. The Junction box is connected to the onshore station by a 40 km electro-optical cable.

## III. EXPECTED PERFORMANCES

The ANTARES scientific program is mainly devoted to the detection of neutrinos of astrophysical origin. Neutrinos can be produced in point-like sources, and detected as an excess of events with respect to the background of atmospheric neutrinos. A diffuse neutrinos flux is also expected from a distribution of sources in the whole sky, giving rise to an excess of high energy events with respect to this background.

The important parameters which characterize a neutrino telescope are the effective area, the angular resolution and the energy resolution. In particular, the neutrino effective area is a crucial parameter because it determines the event rate in a detector. The event rate is calculated as the energy convolution between the neutrino effective area $A_\nu^{eff}$ and the differential flux of neutrinos $d\Phi/dE_\nu$. Figure 2 (upper panel) shows the neutrino effective area $A_\nu^{eff}$ as a function of the neutrino energy, for different values of the neutrino direction. The effective area increases rapidly up to about 1 PeV; above this energy, the value saturates because of the balance between the increase of the hadron-neutrino cross section and the Earth absorption. The Earth absorption strongly decreases the effective area for events below the horizontal. The lower panel of figure 2 shows the muon effective area, which gives the response function of the detector to an incident muon flux, whatever the process that gave rise to this flux. The figure was obtained from simulated charged current neutrino interactions. The square points correspond to reconstructed muons matching the neutrino direction at least by 0.3 degrees.

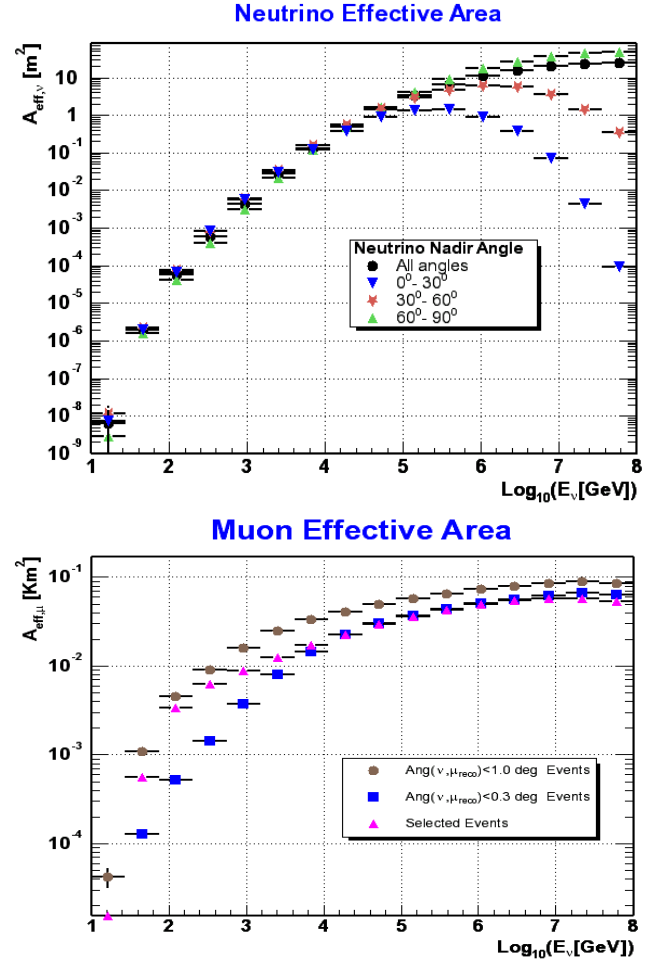

Fig. 2. *Upper panel*. Neutrino effective area vs. neutrino energy for different values of the neutrino direction. The full Monte Carlo chain was used to obtain the distributions; in particular the reconstructed tracks must fulfill some quality cuts. The circles represent the values for a uniform neutrino distribution. *Lower panel*. The muon effective area of the ANTARES telescope. The blue points represent the fraction of events surviving the analysis cuts necessary to have a reconstructed muon direction which differs at most of $0.3^o$ from the parent neutrino direction.

The *intrinsic angular reconstruction* of the telescope is defined as the median angular separation between the true neutrino direction and the reconstructed muon track. The angular resolution is also estimated from simulations, which include all the details of the detector calibration procedure (clock calibration system, LEDs in Optical Modules, optical beacons, laser beacons), described in the next section. Figure 3 shows the median value of the distribution of the angle between the reconstructed and the generated neutrino (circles) or muon (triangles), as a function of the neutrino energy $E_\nu$. At smaller energies the angle between the reconstructed muon and the parent neutrino is dominated by the kinematics effects of the charged current interaction. At $E_\nu >10$ TeV the pointing accuracy is only limited by the intrinsic angular resolution of the detector, which reaches a value of $0.2^o \div 0.3^o$.

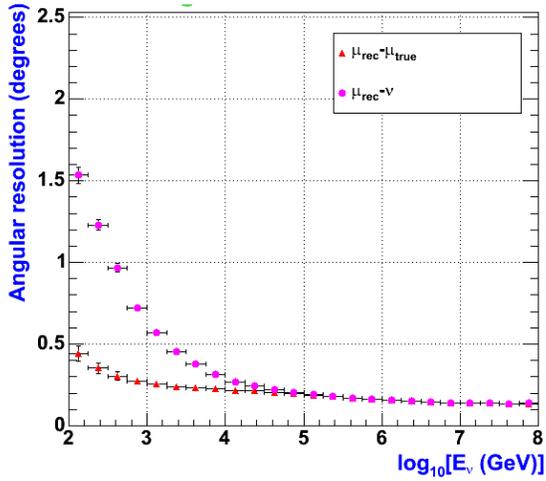

Fig. 3. Median angle of the distribution of the angle between the reconstructed muon and the simulated muon (triangle) or the simulated neutrino (circles) versus the neutrino energy:

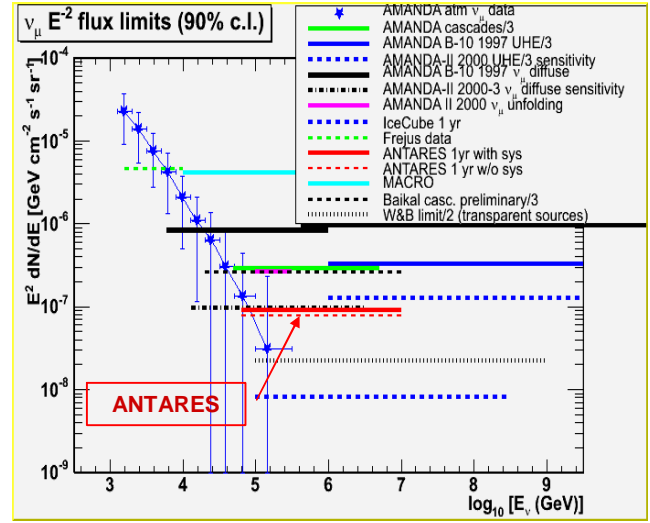

Fig. 4. Experimental limits for different experiments assuming a differential spectrum of neutrinos of $E^{-2}$ spectrum. The sensitivity of some experiments are reported, together with the expected background of atmospheric neutrino and the Waxman-Bahcall limit.

The method of reconstructing the muon energy, the energy resolution, and how to estimate the parent neutrino energy are discussed in [15]. The muon energy is obtained from the measurement of the muon range at small energies, and from the Cherenkov photon intensity at high energies. In fact, due to radiative energy loss processes, the number of Cherenlov photons increases with increasing muon energy. This makes the recorded light quantity an estimator of the muon energy and, thus, of the parent neutrino. The resulting energy resolution is about a factor ~2 above 1 TeV. This allows the definition of selection criteria, in order to suppress the lower energy atmospheric neutrinos and to enhance the high energy ones. The latter are expected from the cumulative contribution of diffuse sources of high energy neutrinos. The energy spectrum from the diffuse sources is expected to be harder (differential spectral index $\gamma \approx 2$) than the energy spectrum of atmospheric neutrinos ($\gamma \approx 3.7$). High energy selected events are composed of an enriched sample of cosmic neutrinos. The ANTARES detector in 1 year of data acquisition will be capable to constrain models (for a $E^{-2}$ flux) predicting muon neutrinos fluxes $d\Phi / dE_\nu \geq 8 \times 10^{-8} E^{-2} GeV^{-1} cm^{-2} s^{-1} sr^{-1}$ [15]. Figure 4 shows the limits for different experiments, compared with the Waixman-Bahcall [16] one.

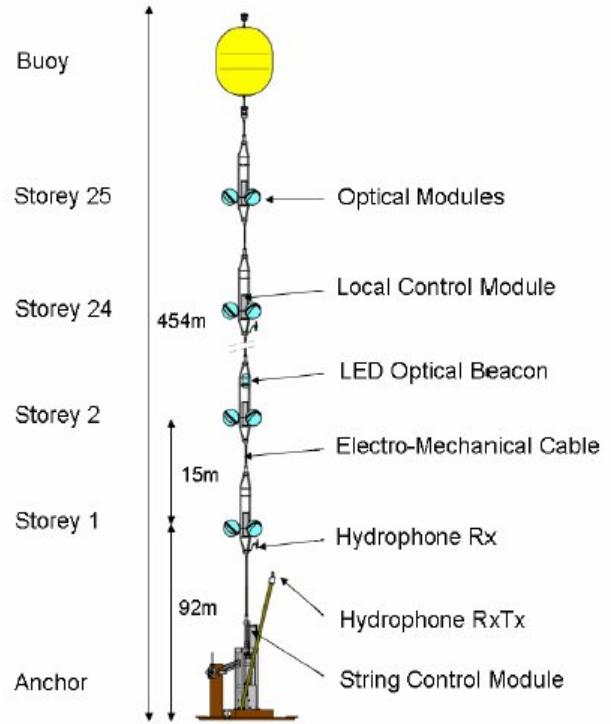

Fig. 5 Schematic view of the first ANTARES detector line. For clarity, only 4 out of 25 storeys are shown and the electro-optical cable lengths are not drawn to scale.

## IV. CONSTRUCTION AND DEPLOYMENT OF MILOM AND LINE 1

Following many tests carried out over several years, the final mechanical and electronic design was chosen for a prototype line (called MILOM). It was deployed and connected to the Junction Box in March 2005 [17]. The MILOM, which uses the final detector design, includes four optical modules and several calibration/monitoring instruments. With the function of an anchor, the Bottom String

Socket (BSS) contains, in addition to the acoustic release system of the string, a seismometer, an acoustic position transducer and a laser beacon. The first storey, located about 100 m above the BSS, has a LED beacon, an acoustic positioning hydrophone, a light transmission meter and a conductivity/temperature probe. Storey 2 (15 m above storey 1) has three optical modules and a sound velocimeter. Storey 3 (50 m above storey 2) has a hydrophone, one optical module, an acoustic Doppler current profiler and a LED beacon. A detailed description of the stand-alone MILOM data acquisition is in [17].

A major step toward the construction of the full detector was achieved during 2006. The first complete line (called Line 1) was deployed on February, 14$^{th}$ 2006 and connected to the shore on March, 2$^{nd}$ 2006. The second line was deployed on July, 15$^{th}$ 2006 and connected to the shore on September, 21$^{st}$ 2006. The following reports the status and the data collected by Line 1 and MILOM. The distance between lines and the MILOM is around 70 m. The 12 lines of the detector will be essentially the same with only minor differences in the instruments used to monitor the sea environment and figure 5 gives a schematic of the line structure. Line 1, as all others, holds 75 optical modules (OM) each containing a 10" Hamamatsu photo-multiplier with an active base supplying the high voltage. The three OMs on each storey are connected to the electronics container. This Local Control Module (LCM), contains the electronics cards to control the high voltage on the optical modules and to digitize the analogue signals from the photomultipliers. The digitized signals are then sent to shore and recorded following a data selection performed with a farm of computer. At the bottom of each line, there is a Bottom String Socket (BSS), from which each line can be released. The line is maintained in an almost vertical position by a buoy located at the top.

Besides the optical modules, the line contains also the *acoustic positioning* and *optical calibration* systems. A good knowledge of the position of each storey, and a time resolution of the order of 1-2 ns, is mandatory to reach the angular resolution of the telescope as shown in figure 3.

The *acoustic positioning* system consists of a hydrophone transducer, (RxTx, in figure 5), located on the base and five hydrophone receiver (Rx, in figure 5) on storeys 1, 8, 14, 20, 25. In addition on the sea bed there is a transducer on the base of the MILOM line and transponders on two autonomous structures, "pyramids" located respectively 150 and 225 m far from Line 1. The *optical calibration* system consists of an optical beacon timing calibration system, which is intended to verify and, if necessary, correct the time-offsets obtained onshore.

Other calibration devices on the line comprise: a compass/tiltmeter device in the Local Control Module of each storey; humidity and temperature sensors and a sound velocity meter at the base of the line.

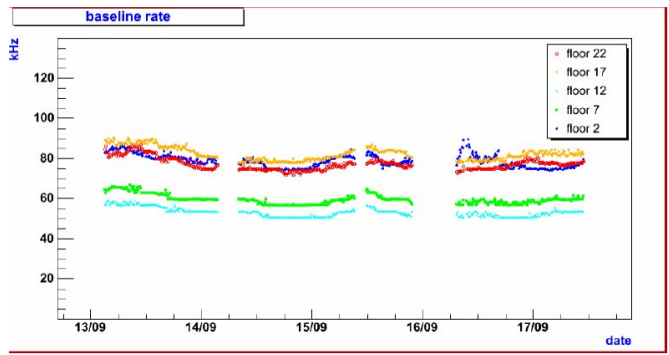

Fig. 6. Counting rates (in kHz) per optical modules. The five colored points (in the electronic version of the paper) show the rate on 5 different storeys (number 2,7,12,17 and 22, numbered from the bottom). The holes represent periods of calibrations, or interruptions in the data taking. We choose, as an example, the week immediately after the end of the 2006 ECRS Conference. The average counting rate increases from the bottom layers (the depth of storey 2 is around 2350 m) to shallower layers (the depth of storey 22 is around 2050 m). During the Spring, the counting rates were higher (from 100 kHz per OM in the bottom, to 500 kHz per OM in the upper detector),

V. LINE 1 AND MILOM PERFORMANCES

A. *Data acquisition*

The ANTARES Optical Module consists of a 10" Hamamatsu photomultiplier tube (PMT) housed in a pressure resistant glass sphere. The 900 PMTs foreseen for the ANTARES detector have been selected to work at a threshold below the single photo-electron level with a mean transit time spread (TTS) of $\sigma \sim 1.3$ ns. The PMT signal is processed by the Analogue Ring Sampler (ARS), an ASIC card which measures the arrival time and charge of the pulse. The readout trigger threshold of the ARS is set at ~0.5 photo-electrons. Figure 6 shows an example of the counting rates recorded by five OMs located on Line 1 for a period of one week. The counting rates exhibit a baseline largely dominated by optical background due to $^{40}$K decays and bioluminescence coming from bacteria, as well as bursts of a few seconds duration, probably produced by bioluminescent emission of macro-organisms [14].

The optical modules deliver their data in real time and can be remotely controlled from the ANTARES shore station through a Gb Ethernet network. Every storey is equipped with a Local Control Module (LCM) which contains the electronics boards for the OM signal processing, the instrument readout, the acoustic positioning, the power system and the data transmission. Every five storeys, in the middle, the Master Local Control Module (MLCM) also contains an Ethernet switch board, which multiplexes the data acquisition (DAQ) channels from the other storeys. At the bottom of the line, the BSS is equipped with a String Control Module (SCM) which contains the local readout and DAQ electronics, as well as the power system for the whole line. Finally, both MCLM and SCM include a Dense Wavelength Division Multiplexing system used for data transmission in order to merge several 1Gb/s Ethernet channels on the same pair of optical fibres, using different laser wavelengths. Although a local trigger requiring time coincidences between OMs of the same storey can be activated, most of the time the large bandwidth of the DAQ system allows the transmission of all recorded OM signals to shore. A dedicated computer farm can then perform a global selection of the OM hits of the interesting physics events from the data recorded by the whole detector.

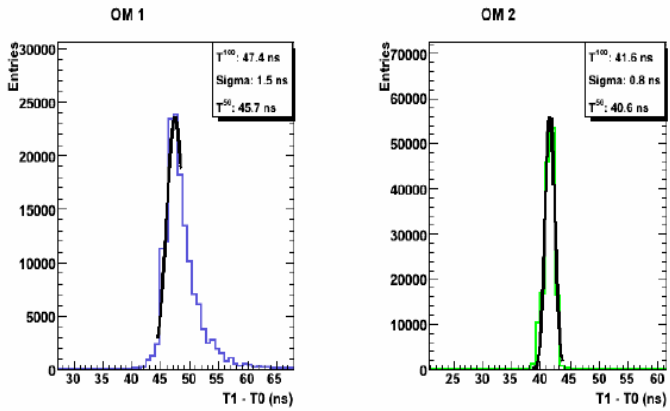

Fig. 7. Data from the optical modules of storey 3 of Line 1 detecting light flashes from the LED optical beacon on the MILOM line. OM2 (right) facing the MILOM see direct light; OM1 (left) hidden on the opposite side sees only scattered light. The distribution from the third OM (which see also direct light) is not shown.

### B. Calibration of Optical Modules

For a detailed timing calibration of the optical modules, a *in-situ* timing calibration system is necessary. This is made of optical beacons. On the detector there are two types of optical beacons: one uses LEDs as light sources, and one uses a laser. There will be four LED beacons along the detector lines, and one on the MILOM line. A laser beacon will be on the base of a small number of lines (at present, only on the MILOM). Each LED beacon contains 36 individual, pulsed LEDs, emitting in the blue with a wavelength spectrum peaked at 470 nm. The beacon emit (roughly) isotropic light with approximately $10^9$ photon/pulse.

Figure 7 shows an example of the data recorded in storey 3 of Line 1, flashing the LED beacon in the MILOM. The OM on Line 1 which face the MILOM (OM2 in the figure) see a narrow distribution dominated by direct light. The OM looking in the opposite direction (OM1 in the figure), only sees scattered light: the measured distribution, with a long tail, is related to the scattering properties of the sea water. For the OMs seeing direct light the distributions are Gaussian with σ ~0.6 – 0.8 ns, as required for the design angular resolution of the detector.

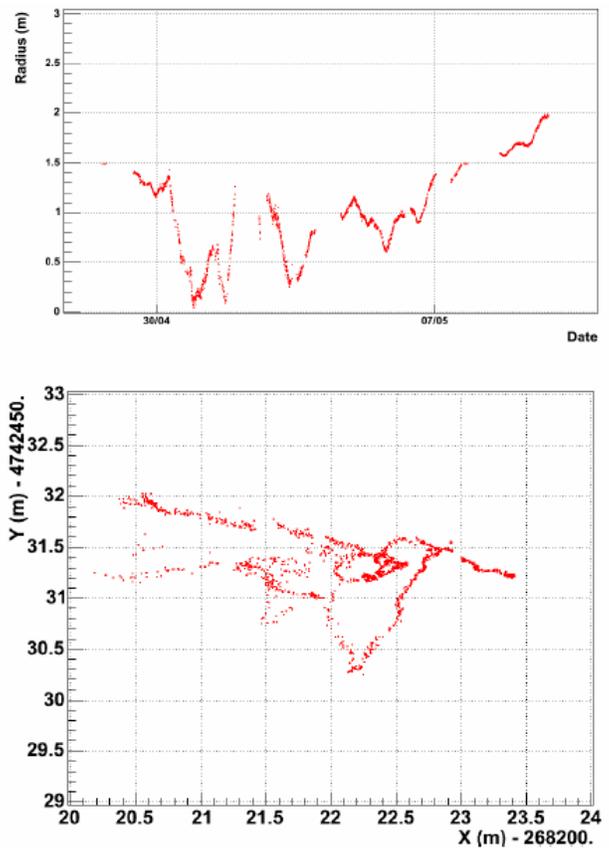

Fig. 8. Example of 12 days of data from the acoustic positioning system for the hydrophone position on storey 1 of Line 1. Top: radial displacement in horizontal plane. Bottom: displacement in X-Y in horizontal plane (with respect to the fiducial position of the line).

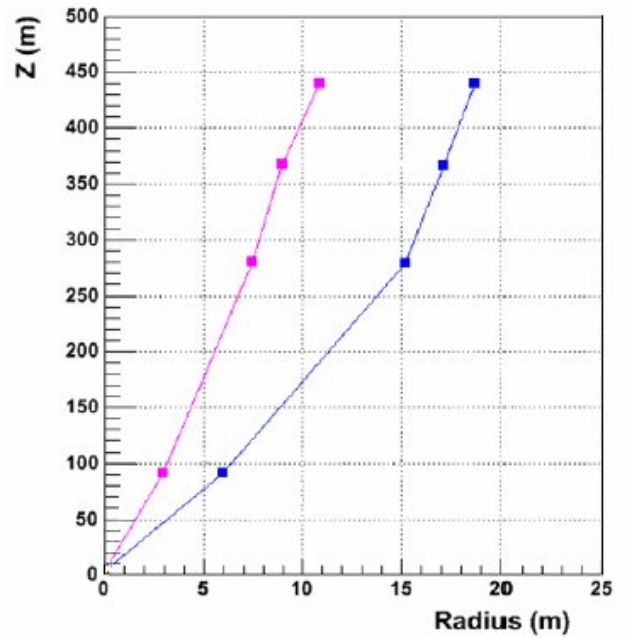

Fig. 9 Reconstructed shape of the line from measurement of the horizontal displacement of four hydrophones on Line 1. The examples are chosen from left to right to correspond to times when the typical current speed was ~12 and ~26 cm/s.

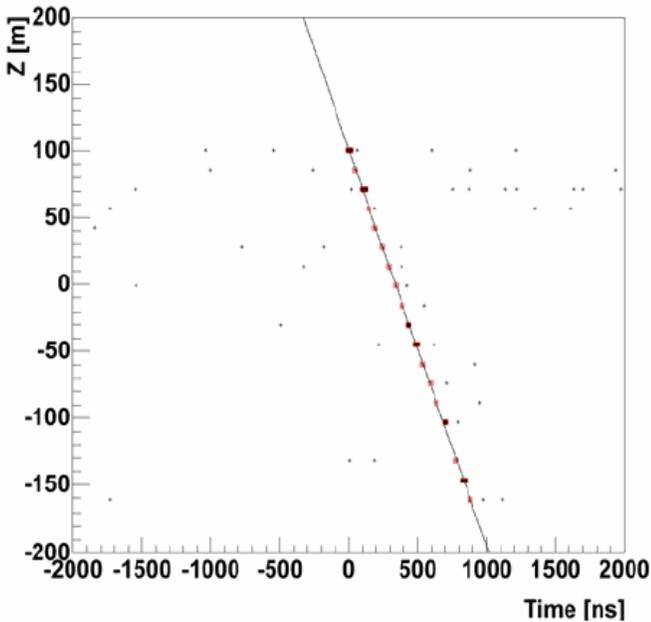

Fig. 10 An atmospheric muon event recorded by Line 1. The display shows the measured hits in the OMs at different heights, Z *vs.* the recorded time (relative to an arbitrary zero). All hits (including background) within a 4 μs time window are displayed as crosses, those in the event trigger as dots and those used in the event fit are surrounded by a red square.

### C. Acoustic positioning data

The high precision acoustic positioning system of ANTARES will consist of a network of fixed emitting and receiving transponders on the sea bed and five receiving hydrophones on each of the detector lines. The network consists of a transducer on the base of every line and a number of autonomous transponders on the sea bed structures called "pyramids". For the "Line 1" operations, there is a transducer on the MILOM base, and two pyramids.

The system works by exchanging sound signals, in the 40-60 kHz frequency range, between the various devices and measuring the time delays between the emission and reception. This delay gives the distance separation, if the sound velocity is known. As an example of the triangulated space points, figure 8 indicates the measured positions (horizontal displacement) of the hydrophone on storey #1 of Line 1 during the period 29 April to 10 May 2006. The individual points are given at intervals of ~5 min. The reproducibility of the measurements close in time indicate a statistical precision of the measurement ~1cm. The systematic errors in the system (mainly due to the uncertainty on the sound velocity) are likely larger than the statistical precision. Figure 9 gives examples of the reconstructed line shape at two particular times when the current speed had different values of ~12 and ~26 cm/s. The plot shows the horizontal displacements for the hydrophones on storeys 1, 14, 20 and 25. The signal intensity received from the hydrophone on storey 8 was not adequate.

### D. Muon reconstruction

ANTARES is designed to observe upward going muons from neutrinos which cross the Earth. However, the angular acceptance of the photomultipliers is such that downward going atmospheric muon can be observed and reconstructed as well. Atmospheric muons are $\sim 10^6$ more abundant than atmospheric neutrinos. They represent a good tool in order to calibrate the detector, but represent also a possible background [18]. As an example, a deficit of atmospheric muons in correspondence to the moon position in the sky can be used to validate the knowledge of the detector absolute positioning.

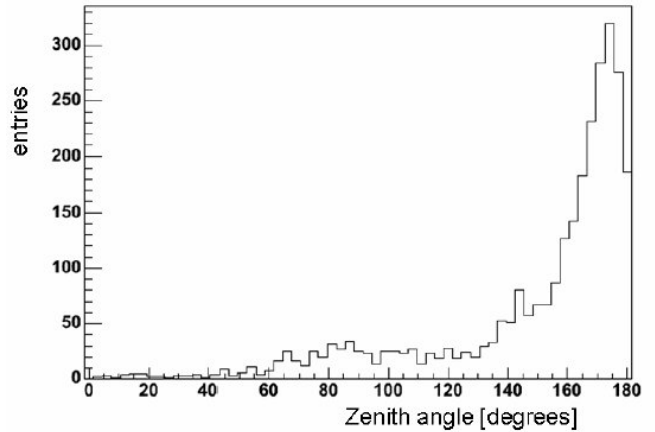

Fig. 11 Zenith angle distribution of reconstructed muon tracks for a subset of events collected during the Line 1 operation.

The trigger condition applied for Line 1 operations by the online computer farm requires at least 5 pairs of OMs with signals coincident within 20 ns. The 5 signal pairs must be causally connected with a muon propagating with the light speeds [19]. Figure 10 shows a display of a typical muon event together with their reconstructed fit.

The reconstruction with only one line suffers from geometrical ambiguity and reduced angular resolution with respect to the full detector. The zenith angle distribution of the reconstructed muon tracks, using a reduced subset of data from Line 1 is shown in figure 11. This distribution of downward going cosmic ray muons (at a zenith angle ~ 180°) is affected by a peak at ~90° due to "ghost events". For these events, the Cherenkov cone is only partially observed, resulting in an ambiguous time pattern with respect to the true track at twice the Cherenkov angle (which is ~43° in water). Such an ambiguity is inherent for a single line reconstruction, but can be minimized by various geometrical cuts. With more than one line this ambiguity will be fully resolved. Work is in progress to compare the muon reconstruction data with Monte-Carlo simulations. To improve the angular resolution, it is necessary to make precise timing corrections using the calibrations. In addition, the *real time* position of the OM must be used, using the corrections from the line shape measures from the acoustic data system.

## VI. SUMMARY AND CONCLUSION

The first complete detector line of the ANTARES neutrino telescope has been constructed, deployed in the sea, connected and at present is in data taking. The data from the first months of operation of the Line 1 and MILOM have demonstrated the stability of the photomultipliers and electronics[*]. The acoustic positioning system has been tested, proving the ability to reconstruct in space the position of the complete line. The optical calibration systems were used to measure the PMTs time offsets. These operations were necessary to reach the quoted angular resolution of the telescope used in the Monte Carlo simulations. Downward going cosmic rays muons have been reconstructed, demonstrating the quality of the reconstruction and analysis software. The counting rates on all the 75 OM were continuously monitored, starting from February. As already observed during the operation with the MILOM in 2005 [17], during March, April and May the counting rates were very high due to bioluminescent organisms. During these months, the data acquisition was not possible with full efficiency. In the remaining part of the year, the rates were of the order of ~100 kHz in most of the detector PMTs. The correlations between bioluminescent bursts (during which the detector is blind) and water parameters (as for example the submarine currents) were in study. Figure 12 shows the correlation between the percentage of burst fraction during the acquisition time, and the water current speed. The current speed is measured in the MILOM with an acoustic direct reading ADCP current profiler [20].

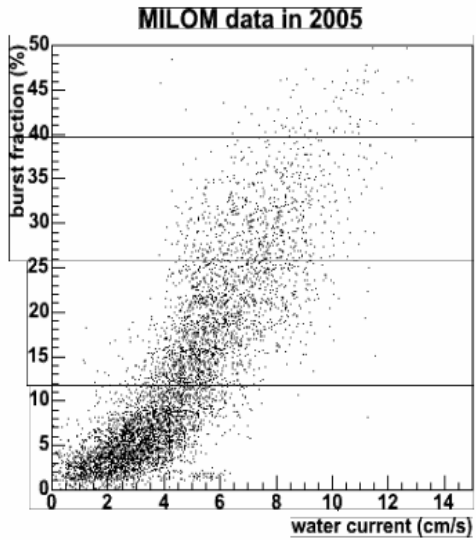

Fig. 11  Correlation between the percentage of data taking with burst fraction and the submarine current speed for one of the OM of the MILOM [21].

The successful operation of Line 1 proves the feasibility of the technology chosen for the undersea neutrino detector. The long-term running of the main electro-optical cable and junction box on the seabed for more than four years demonstrates the reliability of this vital part of the system which is the only single-point failure element in the detector design.

Further, the performance of the MILOM line, with no failures after deployment, for more than one and ½ year in the sea shows that the electronics systems are reliable, albeit with low numbers of modules.

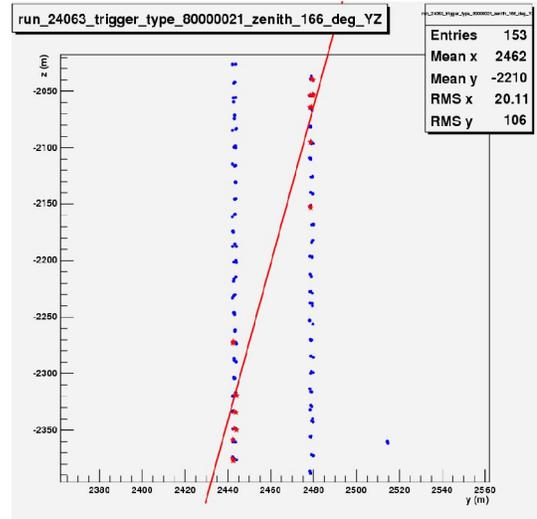

Fig. 12  One of the first muons seen with two strings. The Line 2 was connected in 2006 September, 21th The plot shows (in reds) the hits which triggered the events, and which were used for the reconstruction of the track (red line).

Given these achievements, the production of components and assembly of the remaining ANTARES lines is now in progress. The second line was already deployed and connected, and the first events already reconstructed (see figure 12). We plans to complete the detector by the end of 2007.

[*] Line 2 was connected when this paper was in preparation

**APPENDIX**. THE ANTARES COLLABORATION

J.A. Aguilar[j], A. Albert[u], F. Ameli[x], M. Anghinolfi[i],G. Anton[g], S. Anvar[y], E. Aslanides[e], J-J. Aubert[e], E. Barbarito[b], S. Basa[r], M. Battaglieri[i], Y. Becherini[c], R. Bellotti[b], J. Beltramelli[y], V. Bertin[e], A. Bigi[w], M. Billault[e], R. Blaes[u], N. de Botton[y], M.C. Bouwhuis[v], S.M. Bradbury[t], R. Bruijn[v,ab], J. Brunner[e], G.F. Burgio[f], J. Busto[e], F. Cafagna[b], L. Caillat[e], A. Calzas[e], A. Capone[x], L. Caponetto[f], E. Carmona[j], J. Carr[e], S.L. Cartwright[z], D. Castel[u], E. Castorina[w], C. Cavasinni[w], S. Cecchini[c,m], A. Ceres[b], P. Charvis[h], P. Chauchot[k], T. Chiarusi[x], M. Circella[b], C. Colnard[v], C. Compère[k], R. Coniglione[s], N. Cottini[w], P. Coyle[e], S. Cuneo[i], A-S. Cussatlegras[d], G. Damy[k], R. van Dantzig[v], C. De Marzo[b1], I. Dekeyser[d], E. Delagnes[y], D. Denans[y], A. Deschamps[h], F. Dessages-Ardellier[y], J-J. Destelle[e], B. Dinkespieler[e], C. Distefano[s], C. Donzaud[y], J-F. Drogou[ℓ], F. Druillole[y], D. Durand[y], J-P. Ernenwein[u], S. Escoffier[e], E. Falchini[w], S. Favard[e], F. Feinstein[e], S. Ferry[n], D. Festy[k], C. Fiorello[b], V. Flaminio[w], S. Galeotti[w], J-M. Gallone[n], G. Giacomelli[c], N. Girard[u], C. Gojak[e], Ph. Goret[y], K. Graf[g], G. Hallewell[e], M.N. Harakeh[q], B. Hartmann[g], A. Heijboer[v,ab], E. Heine[v], Y. Hello[h], J.J. Hernández-Rey[j], J. Hößl[g], C. Hoffman[n], J. Hogenbirk[v], J.R. Hubbard[y], M. Jaquet[e], M. Jaspers[v,ab], M. de Jong[v], F. Jouvenot[y], N. Kalantar-Nayestanaki[q], A. Kappes[g], T. Karg[g], S. Karkar[e], U. Katz[g], P. Keller[e], H. Kok[v], P. Kooijman[v,aa], C. Kopper[g], E.V. Korolkova[z], A. Kouchner[a], W. Kretschmer[g], A. Kruijer[v], S. Kuch[g], V.A. Kudryavtsev[z], D. Lachartre[y], H. Lafoux[y], P. Lagier[e], R. Lahmann[g], G. Lamanna[e], P. Lamare[y], J.C. Languillat[y], H. Laschinsky[g], Y. Le Guen[k], H. Le Provost[y], A. Le Van Suu[e], T. Legou[e], G. Lim[v,ab], L. Lo Nigro[f], D. Lo Presti[f], H. Loehner[q], S. Loucatos[y], F. Louis[y], F. Lucarelli[x], V. Lyashuk[p], M. Marcelin[r], A. Margiotta[c], R. Masullo[x], F. Maz´eas[k], A. Mazure[r], J.E. McMillan[z], R. Megna[b], M. Melissas[e], E. Migneco[s], A. Milovanovic[t], M. Mongelli[b], T. Montaruli[b], M. Morganti[w], L. Moscoso[y,a], M. Musumeci[s], C. Naumann[g], M. Naumann-Godo[g], V. Niess[e], C. Olivetto[n], R. Ostasch[g], N. Palanque-Delabrouille[y], P. Payre[e], H. Peek[v], C. Petta[f], P. Piattelli[s], J-P. Pineau[n], J. Poinsignon[y], V. Popa[c,o], T. Pradier[n], C. Racca[n], N. Randazzo[f] , J. van Randwijk[v], D. Real[j], B. van Rens[v], F. Réthoré[e], P. Rewiersma[v1], G. Riccobene[s], V. Rigaud[ℓ], M. Ripani[i], V. Roca[j], C. Roda[w], J.F. Rolin[k], M. Romita[b], H.J. Rose[t], A. Rostovtsev[p], J. Roux[e], M. Ruppi[b], G.V. Russo[f], F. Salesa[j], K. Salomon[g], P. Sapienza[s], F. Schmitt[g], J-P. Schuller[x], R. Shadnize[g], I. Sokalski[b], T. Spona[g], M. Spurio[c], G. van der Steenhoven[v], T. Stolarczyk[y], K. Streeb[g], D. Stubert[u], L. Sulak[e], M. Taiuti[i], C. Tamburini[d], C. Tao[e], G. Terreni[w], L.F. Thompson[z], P. Valdy[ℓ], V. Valente[x], B. Vallage[y], G. Venekamp[v], B. Verlaat[v], P. Vernin[y], R. de Vita[i], G. de Vries[v,aa], R. van Wijk[v], P. de Witt Huberts[v], G. Wobbe[g], E. de Wolf[v,ab], A-F. Yao[d], D. Zaborov[p], H. Zaccone[y], J.D. Zornoza[j], J. Zúñiga[j]

1 Deceased.
a APC – AstroParticule et Cosmologie, Universitè Paris 7, France
b Dipartimento Interateneo di Fisica e Sezione INFN, Bari, Italy
c Dip. Fisica dell'Università e Sezione INFN, Bologna, Italy
d COM – Centre d'Ocèanologie de Marseille, France
e CPPM – Centre de Physique des Particules de Marseille, France
f Dip. Fisica ed Astronomia dell'Università e Sezione INFN, Catania, Italy
g Friedrich-Alexander-Universität Erlangen-Nürnberg, Germany
h GéoSciences Azur, CNRS/INSU, Villefranche-sur-Mer, France
i Dip.Fisica dell'Università e Sezione INFN, Genova, Italy
j IFIC – Instituto de Física Corpuscular,  Valènica, spain
k IFREMER – Centre de Brest, France
ℓ IFREMER – Centre de Toulon/La Seyne Sur Mer, France
m INAF-IASF, Bologna, Italy
n IPHC, Université Louis Pasteur et IN2P3/CNRS, Strasbourg 23, France
o Institute for Space Sciences, 77125 Bucharest, Magurele, Romania
p ITEP – Institute for Theoretical and Exp. Physics, Moscow, Russia
q KVI, University of Groningen, The Netherlands
r LAM, CNRS/INSU et Université de Provence, Marseille, France
s INFN – LNS, Catania, Italy
t School of Physics & Astronomy, University of Leeds LS2 9JT, UK
u GRPHE Université de Haute Alsace, Mulhouse Cedex, France
v NIKHEF, Amsterdam, The Netherlands
w Dip. Fisica  dell'Università e Sezione INFN, Pisa, Italy
x  Dip. Fisica "La Sapienza" e Sezione INFN,  Roma, Italy
y DSM/Dapnia Gif-sur-Yvette Cedex, France
z Dept. of Physics and Astronomy, University of Sheffield, UK
aa Universiteit Utrecht, Utrecht, The Netherlands
ab Universiteit van Amsterdam,  Amsterdam, The Netherlands